\documentstyle[epsfig]{aipproc}

\begin{document}
\title{Broad Band X-ray Spectral Properties\\ of Gamma-Ray
Bursts with BeppoSAX}

\author {F. Frontera$^{a,b}$, L. Amati$^{c,f}$, E. Costa$^{c}$, M. Feroci$^{c}$, 
J.M. Muller$^{d,e}$,
G. Pizzichini$^{a}$, M.N. Cinti$^{c}$, D. Dal Fiume$^{a}$, J. Heise$^{d}$, L. Nicastro$^{a}$, 
M. Orlandini$^{a}$,
E. Palazzi$^{a}$, J. in 't Zand$^{d}$}

\address{
$^{a}$Istituto TESRE, CNR, Bologna, Italy \\
$^{b}$Dipartimento di Fisica, Universit\`a, Ferrara, Italy \\
$^{c}$Istituto di Astrofisica Spaziale, CNR, Frascati (RM), Italy \\
$^{d}$Space Research Organization in the Netherlands, Utrecht, The Netherlands \\
$^{e}$BeppoSAX Scientific Data Center, Rome, Italy \\
$^{f}$Istituto Astronomico, Universit\`a La Sapienza, Roma, Italy}

\maketitle

\begin{abstract}
In about one year, five gamma--ray bursts were simultaneously observed with the
Wide Field Cameras and Gamma Ray Burst Monitor aboard the BeppoSAX satellite.
From some of them X--ray afterglow emission has been clearly detected with the
 same satellite. In order to understand how GRB emission is related to the
 X--ray afterglow, we are performing a systematic study of the spectral 
properties of these events. We report here preliminary results of this study.
\end{abstract}

\section*{Introduction}

The discovery by BeppoSAX \cite{Boella97a} of the first afterglow sources of
celestial Gamma-Ray Bursts (GRB) has strongly increased the  astrophysical
interest
\begin{table}
\caption{GRBs simultaneously detected from BeppoSAX GRBM and WFCs}
\label{tab:calsources}
\begin{center}
\begin{tabular}{|c|c|c|c|c|} \hline
\rule{0pt}{2.5ex}  GRB  & $\gamma$-ray Fluence  & $\gamma$-ray Duration & 
TOO time delay & 
X-ray afterglow \\
  & ($10^{-5}$erg/cm$^{2}$) & (sec) &  & (yes/no)  \\[0.2ex] \hline
   GRB960720 & $0.34\pm{0.40}$ & 8 & $43^{d}$ & ? \cite{Piro96}\\ \hline
   GRB970111 & $4.14\pm{0.30}$ & 43 & $16^{h}$ & ? \cite{feroci97a}\\ \hline
   GRB970228 & $1.17\pm{0.20}$ & 55 & $8^{h}$  & yes \cite{costa}\\ \hline
   GRB970402 & $0.91\pm{0.09}$ & 150 & $8^{h}$  & yes \cite{nicastro}\\ \hline
   GRB970508 & $0.18\pm{0.03}$ & 15 & $5.7^{h}$  & yes \cite{Piro97c}\\
\hline
\end{tabular}
\end{center}
\end{table}

\begin{figure}[h!] 
\centerline{\epsfig{file=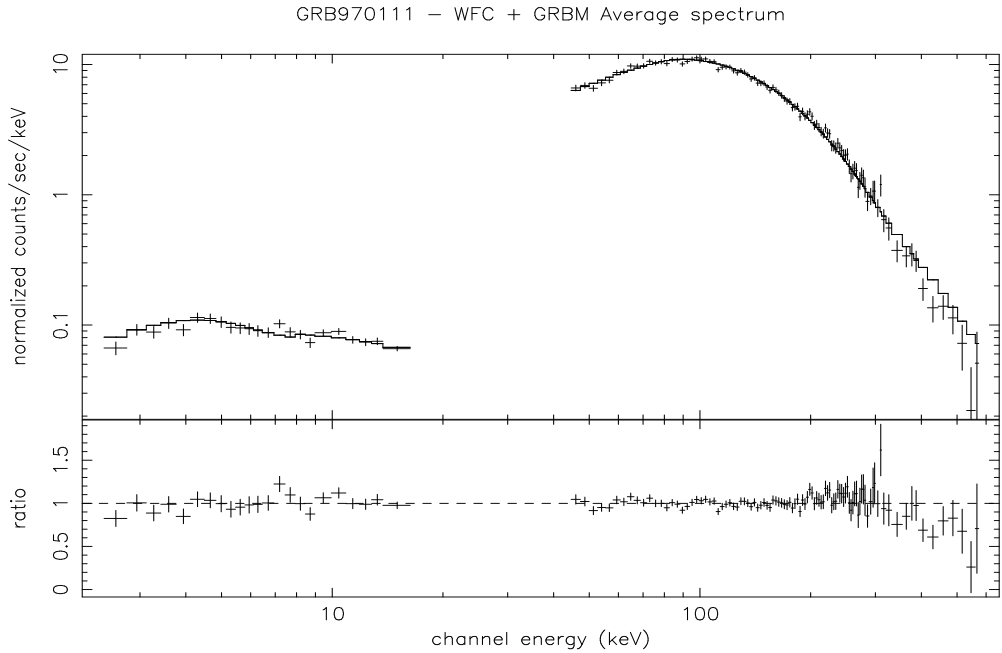,height=6.0in,width=4.0in,
 angle=-90}}
\vspace{10pt}
\caption{WFC/GRBM Time averaged spectrum of GRB970111. The best
fit in the 1.5 to 700 keV energy band is obtained with a broken power law
(see continuous line) with the following parameters:
 $\alpha_1=0.50\pm0.04$,  $\alpha_2=2.13\pm0.03$, $E_{break}
 =101\pm1$ keV, ($\chi^2/195=1.24$).
In the fit the instrument relative normalization was a free parameter.}
\label{fig1}
\end{figure}

for the GRB phenomenon. BeppoSAX offers the possibility not only to
discover the presence of very faint  X-ray afterglow emission
($\geq$5 $\times$ 10$^{-14} erg/(cm^2\,s)$) following a GRB event, but also to
study in a broad energy band (1.5--700~keV) the specific bursts from which
afterglow emission can be searched for. This is possible thanks to the
presence aboard the same satellite of a Gamma-Ray Burst Monitor (GRBM)
\cite{Frontera97a}
and two Wide Field Cameras (WFC)\cite{Jager97}. These instruments have a 
field of view
partially superposed. When an event is simultaneously detected by 
both instruments, the position of the burst can be precisely determined
(within a few arcmin error radius) and thus a follow-on observation of the
GRB error box can be started in other bands of the electromagnetic spectrum
(X-rays, optical, IR, radio).
\\

\begin{figure}[h] 
\centerline{\epsfig{file=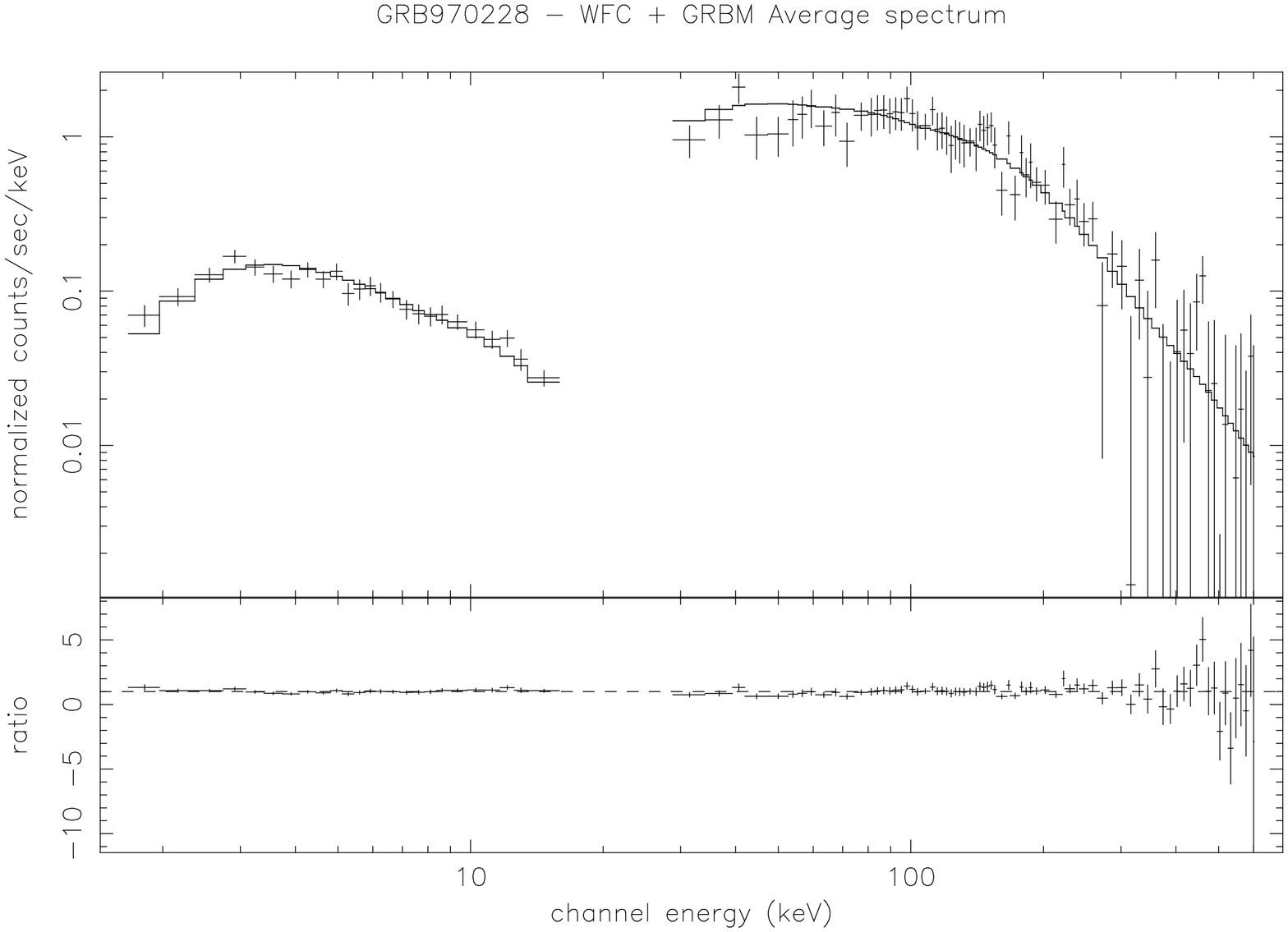,height=6.0in,width=4.0in,
 angle=-90}}
\vspace{10pt}
\caption{WFC/GRBM Time averaged spectrum of GRB970228. The best
fit in the 1.5 to 700~keV energy band is obtained 
with a broken power law (see continuous line) with the following parameters:
 $\alpha_1\,=\,1.35\pm0.07$,  $\alpha_2\,=\,1.95\pm0.05$, $E_{break}\,
=\,13\pm3$ keV, $\chi^2/195\,=\,1.05$ }
\label{fig2}
\end{figure}

\begin{figure}[h] 
\centerline{\epsfig{file=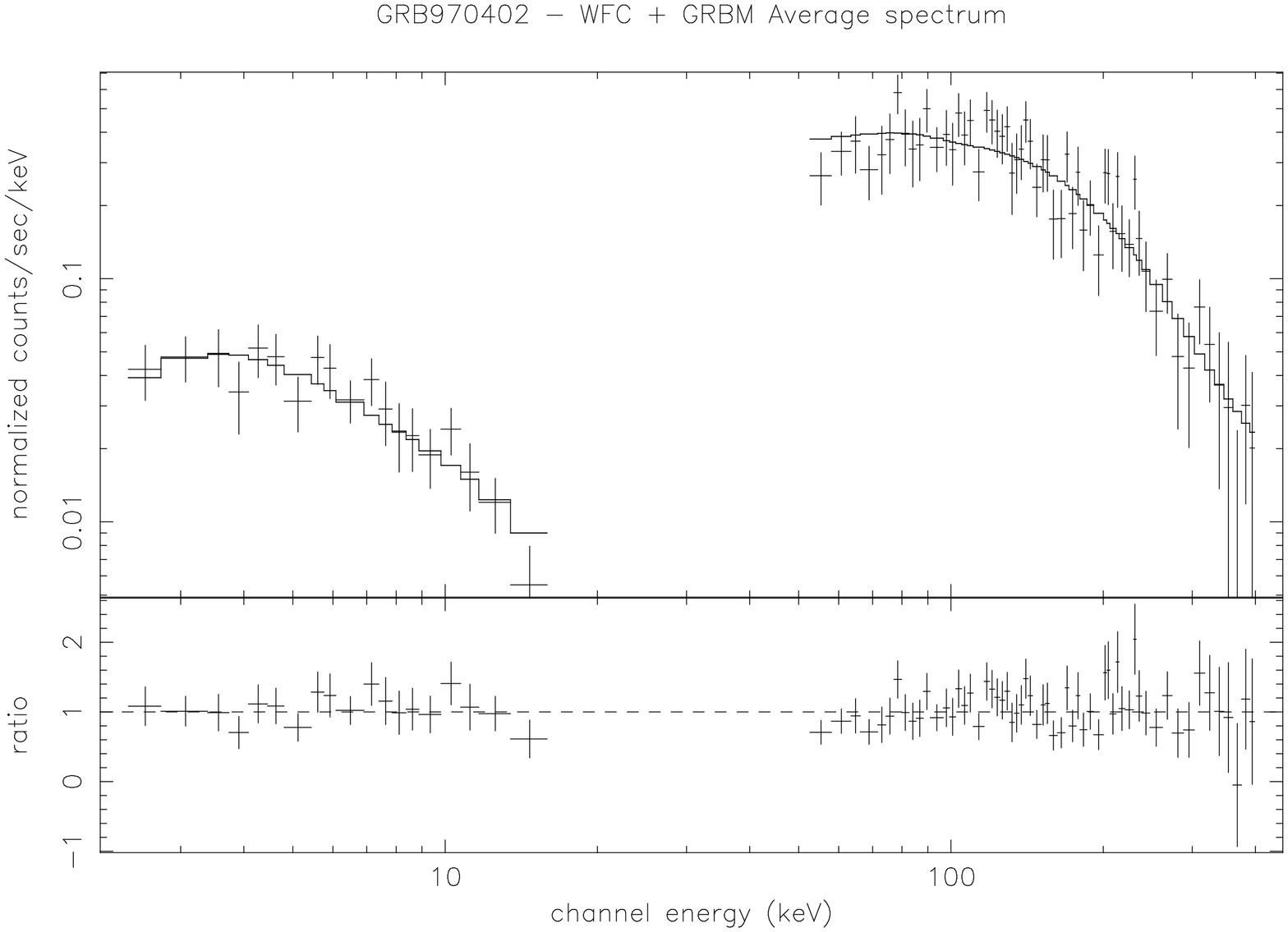,height=6.0in,width=4.0in,
 angle=-90}}
\vspace{10pt}
\caption{WFC/GRBM Time averaged spectrum of GRB970402. The best
fit in the 1.5 to 700~keV energy band is obtained with a single power law
(see continuous line) with  photon index $\alpha\,=\,1.35\pm0.08$
($\chi^2/199 \,=\,0.96$).
}
\label{fig3}
\end{figure}

So far 5 GRBs were simultaneously detected with
GRBM and WFC (see Table 1).
Of them GRB970111 was the strongest event, in terms of peak flux and fluence,
and GRB970402 was the weakest. While search for afterglow 
X-ray emission
from GRB960720 started  after a long time (43 days) from the initial
event\cite{Piro96}, this search for the other bursts  started after about
 6--16~hrs (see Table 1). The result
is known: three clear X-ray afterglow sources, associated with GRB970228
\cite{Costa97a,costa}, GRB970402\cite{Piro97a,nicastro}, GRB970508\cite{Piro97b,Piro97c}, 
respectively, were discovered. For the first and the last of these 
events also an optical afterglow was observed\cite{Groot97,Bond97}. 
In addition, for the last
event (GRB970508) also a radio counterpart was detected\cite{Frail97}.
\\
For the strongest event, GRB970111, there is some evidence of
X--ray afterglow \cite{feroci97a} and no detection of optical or radio 
counterpart. 
It is of primary importance  to understand the reason for that.
\\
A possible signature of the peculiarity of GRB970111 with respect to the
other bursts is its energy spectrum.
Thus a comparison of time averaged energy spectra of these bursts 
can help to solve the problem.

Here we report preliminary results of this spectral analysis.
\section*{RESULTS}
Spectra averaged on time profiles of GRB970111, GRB970228 and GRB970402 
in the broad
energy band from 1.5 to 700 keV were obtained by using both WFC and GRBM
spectral data. A response matrix was obtained for each of these instruments
for on-axis incident photons. This can introduce some systematic errors
in the flux estimate, but not in the spectral shape, as confirmed by the
Crab Nebula spectrum determination \cite{feroci97b}. 
XSPEC software package (v. 9.0) was used
to derive spectral parameters and their uncertainties (1 $\sigma$).
Figures 1, 2 and 3 show preliminary results.

\section*{DISCUSSION}
From the fit results, both GRB970111 and
GRB970228 spectra are fit with a broken power law, while the spectrum of
GRB970402 is fit with a single power law. By comparing the break energy of
the GRB970111 spectrum with that of the GRB970228 spectrum, we see that it
is much higher for GRB97011 than for GRB970228: 101~keV vs. 13~keV.
This fact could be a hint that the peak energy of the $\nu F_\nu$ spectrum
evolved much more rapidly toward lower energies in the case of GRB970228.
A fast evolution of $\nu F_\nu$ was actually observed in GRB970228  
\cite{Frontera97b}. How this different evolution can influence the
presence or not of an afterglow emission is not clear.
\\
Work is in progress to complete the comparative spectral analysis by including
the spectral evolution of the bursts.


\begin{references}
\bibitem{Boella97a}
Boella, G. et~al. 1997a, A\&A Suppl.Ser.,  122, 299.
%
\bibitem{Frontera97a}
Frontera, F., Costa, E., Dal~Fiume, D., Feroci, M., Nicastro, L., Orlandini,
  M., Palazzi, E., and Zavattini, G. 1997a, A\&A Suppl. Ser., 122, 357.
%
\bibitem{Jager97}
Jager, R., et~al. 1997, A\&A Suppl. Ser., in  press.
%
\bibitem{Piro96}
Piro, L. et~al. 1996, IAU Circ. 6480.
%
\bibitem{feroci97a}
Feroci, M. et al. 1997, in preparation
%
\bibitem{costa}
Costa, E. et~al. 1997, Nature, 387, 783
%
\bibitem{nicastro}
Nicastro, L. et al. 1997, in preparation
%
\bibitem{Piro97c}
Piro, L. et~al. 1997c, A\&AL, submitted.
%
\bibitem{Costa97a}
Costa, E. et~al. 1997, IAU Circ. 6576.
%
\bibitem{Piro97a}
Piro, L. et~al. 1997a, IAU Circ. 6617.
%
\bibitem{Piro97b}
Piro, L. et~al. 1997b, IAU Circ. 6656.
%
\bibitem{Groot97}
Groot, P.J. et al. 1997, IAU Circ. 6584.
%
\bibitem{Bond97}
Bond, H. E. 1997, IAU Circ. 6654.
%
\bibitem{Frail97}
Frail, D. A. et al. 1997, IAU Circ. 6662.
%
\bibitem{feroci97b}
Feroci, M. et al. 1997, this proceedings 
%
\bibitem{Frontera97b}
Frontera, F. et~al. 1997, ApJ Lett., accepted.

\end{references}
\end{document}